\documentclass[proceedings, preprint]{rmaa}

\usepackage{paralist}

\usepackage{psfrag,color}

\SetYear{2022}
\SetConfTitle{Prospects for low-frequency radio astronomy in S. A.}

\title{Search for extraplanar radio emission in a sample of edge-on starburst galaxies} 

\author{
  C. A. Galante,\altaffilmark{1,2} 
  J. Saponara,\altaffilmark{2}
  G. E. Romero,\altaffilmark{1,2}
  and P. Benaglia\altaffilmark{2}}

\altaffiltext{1}{Facultad de Ciencias Astronómicas y Geofísicas,
  UNLP, Paseo del Bosque S/N, 1900, La Plata,
  Buenos Aires, Argentina.}

\altaffiltext{2}{Instituto Argentino de Radioastronomía,
  Berazategui, Buenos Aires, Argentina.} 

\shortauthor{Galante et al.}
\shorttitle{Extraplanar radio emission in starburst galaxies}

\listofauthors{C. A. Galante, J. Saponara, G. E. Romero, \& P. Benaglia}
\indexauthor{Galante, C. A.}
\indexauthor{Saponara, J.}
\indexauthor{Romero, G. E.}
\indexauthor{Benaglia, P.}

\abstract{The intense star formation in starburst galaxies results in the formation of intense winds that sweep matter up to several kpc out of the galactic plane. These winds can be detected in many bands of the electromagnetic spectrum.
In this opportunity, we present our project to study three edge-on galaxies, candidates to show extraplanar emission at radio wavelengths. We will describe the criteria followed in selecting the sample and the software developed to detect the presence of the wind. Besides, we will discuss the next steps of this project.}

\resumen{La intensa formación estelar que presentan las galaxias \textit{starburst} resulta en la formación de fuertes vientos que arrastran materia hasta varios kpc por fuera del plano galáctico, emitiendo radiación en varias bandas del espectro electromagnético. En esta oportunidad, presentamos nuestro proyecto para estudiar tres galaxias vistas de canto, candidatas a presentar emisión por fuera del plano galáctico en ondas de radio. Describiremos el criterio utilizado para selecionar la muestra de galaxias y el software creado para detectar la presencia de los vientos. Además, discutiremos los próximos pasos a seguir de este proyecto.}

\addkeyword{Galaxies: halos}
\addkeyword{Galaxies: starburst}
\addkeyword{Radio continuum: galaxies}
\addkeyword{Radiation mechanisms: synchrotron}

\begin{document}
\maketitle

\section{General}
\label{sec:intro}

Starburst galaxies exhibit a remarkable level of star formation activity. The combined effect of intense stellar winds generated by hot stars and a high rate of supernova explosions leads to gas ejection from the star-forming region into the surrounding environment, creating powerful winds \citep{Chevalier}. These winds play an important role in shaping the formation and evolution of galaxies. They can inhibit star formation by depleting the central gas reservoirs \citep{Veilleux}. As the gas escapes, star formation decreases and the outflow stops. However, the arrival of fresh gas can reignite the starburst. This influx of fresh gas is often triggered by gravitational interactions with neighboring galaxies, or when previously ejected gas that remains gravitationally bound falls back into the galaxy as high-speed clouds \citep{delValle}. However, some of the gas may not be able to return to the galactic disk because it is no longer gravitationally bound. Consequently, this gas enriches the intergalactic medium as it is carried away by the outflow, supplementing it with heavy elements.

Different phases make up the outflow, and the relative positions of each phase provide important clues to understanding the three-dimensional picture. The hot gas in the cavity emits X-rays, the diffuse ionized gas surrounding the hot gas emits $\rm H_\alpha$, and the atomic gas in a cooler shell pulled up by a thick disk emits at the 21-cm line \citep[see Fig. 8,][]{Boomsma}. In addition, the presence of magnetic fields and the relativistic particles that emit synchrotron radiation can be studied by detecting the extra-planar radio continuum emission \citep{Krause}. The best example is the starburst galaxy NGC\,253, located at a distance of 3.9~Mpc in the Sculptor group of galaxies. The nuclear starburst is thought to be forming stars at a high rate and producing a superwind that transports material into the halo. The galaxy and its extra-planar gas have been well studied in recent years at various wavelengths \citep{Boomsma,Bauer,heesena,Romero}. \\

We are carrying out a project to detect and characterize the extra-planar non-thermal emission in three edge-on galaxies, to determine the non-thermal contribution, to measure the magnetic field and its distribution, and to detect embedded sources. In the following sections, we describe the sample of galaxies, the method implemented to detect the continuum emission and the current status of this project.

\begin{table*}
  \caption{Sample of galaxies}
  \begin{tabular}{l@{~~}c@{~~}c@{~~}c@{~~}c@{~~}c@{~~}c@{~~}c@{~~}c@{~~}c@{~~}}
\hline
Name     &     RA     &    DEC        &    $i$     &  $D$     &  $\theta$       & $\log(L_{\mathrm{IR}})$   &  $SFR$    & $S_{60}/S_{100}$  &   $S_{1.4}$     \\
         & (hh:mm:ss) &  (dd:mm:ss)   &   (deg)  & (Mpc)  &  (arcmin)& ($L_{\sun}$ ) &  ($\mathrm{M_{\sun}~yr^{-1}}$ ) &    &          (mJy)  \\
\hline                                                                                                            
NGC 1055 & 02:41:45   &    00:26:35   &  83.8    & 17.7   &  6.17           &   10.09     &   1.83  &    0.36   &    200.9   \\
NGC 4527 & 12:34:08   &    02:39:13   &  77.3    & 13.3   &  6.03           &   10.42     &   3.92  &    0.48   &    178     \\
NGC 5690 & 14:37:41   &     02:17:27  &  83.3    & 19.14  &  3              &   10.23     &   2.53  &    0.4    &     72     \\
\hline
\label{tab:sample}
\end{tabular}
\end{table*}

\section{The galaxy sample}
\label{sec:sample}

From the Imperial IRAS-FSC Redshift Catalog \citep{Wang}, we selected the galaxies according to the following criteria: 1. They should have an
inclination higher than $75^{\circ}$ to allow for a clear separation of the halo emission from the disk emission. 2. They must fulfill that $\log(L_{\mathrm{IR}}/L_{\sun})>10$ to guarantee the intense star formation activity. 3. The ratio of infrared fluxes at 60 and 100 $\mu$m should be $S_{60}/S_{100} > 0.4$ to increase the chance of presenting disk-halo interactions \citep{Dahlem}. 4. The angular sizes ($\theta$) and distances ($D$) must satisfy $\theta > 3$~arcmin and $D < 20$~Mpc, to ensure their detection at radio wavelengths and to allow us to resolve the emission in z-direction.

The following galaxies constitute the galaxy sample. We list the main properties of the galaxies, $i$ inclination, $D$ distance, $\theta$ angular size, the $L_{\mathrm{IR}}$, SFR, $S_{60}/S_{100}$ and the flux density at 1.4~GHz ($S_{1.4}$) in Table~\ref{tab:sample}.\\

 NGC\,1055: This galaxy shows evidence of a minor merger observed at the low-angular resolution of the VLA and in Chandra X-ray observations (with low integration time).

NGC\,4527: This galaxy was detected in X-rays by Chandra. Low-angular resolution observations carried out with the VLA suggest the presence of a point-like source embedded in the diffuse radio continuum emission.

NGC\,1421: This galaxy was previously observed with the VLA. \citet{Irwin} suggest evidence of extra-planar emission and possible embedded sources.

\section{Methodology}
\label{sec:meth}

\begin{figure}
    \centering
    \includegraphics[width=0.4\textwidth]{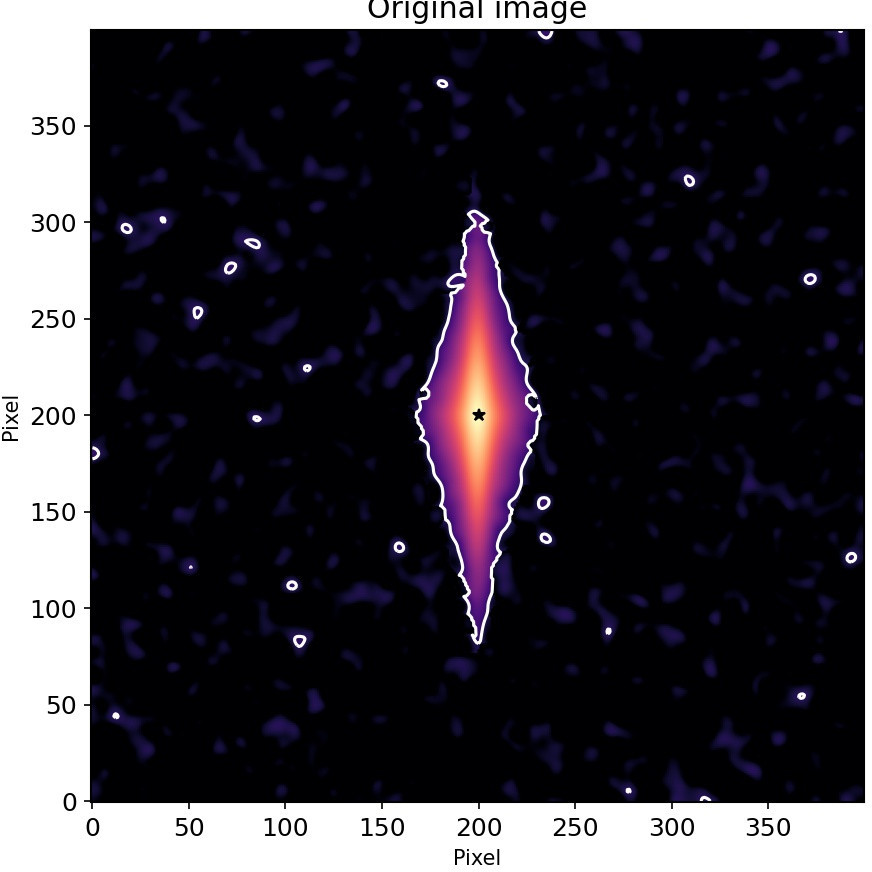}\\
    \includegraphics[width=0.4\textwidth]{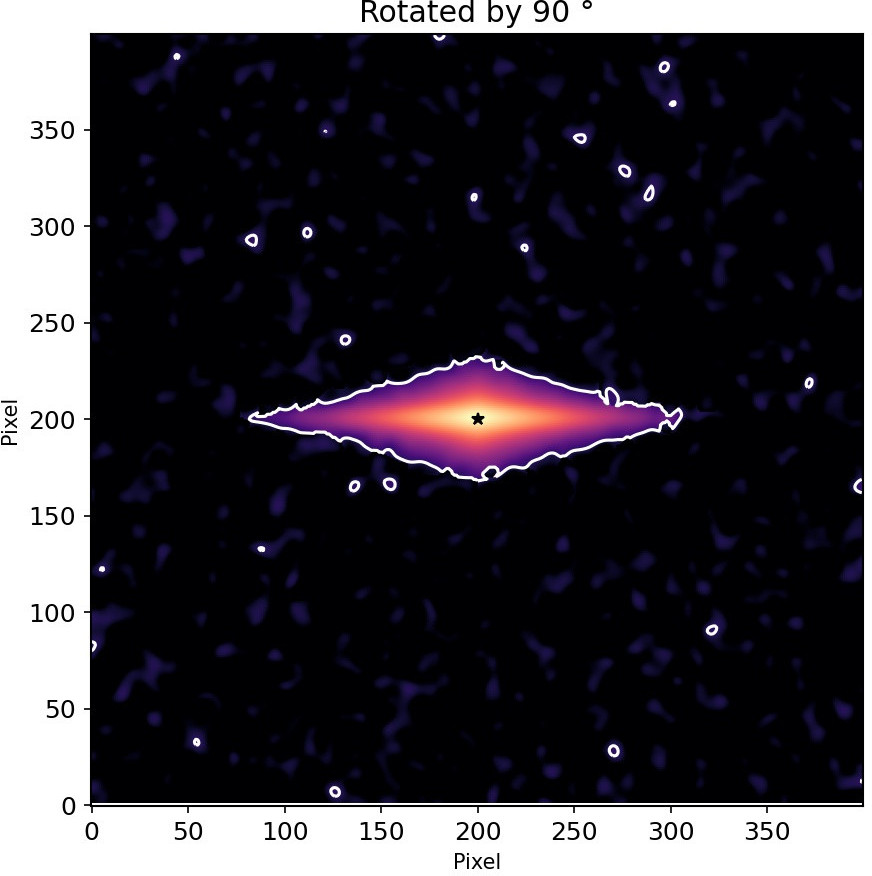}\\
    \includegraphics[width=0.5\textwidth]{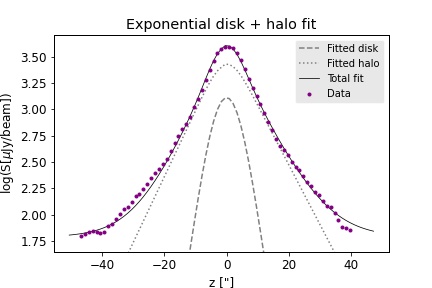}

    \caption{Artificial image of a galaxy and its $z$-profile. \textit{Top:} The artificial image (\textit{upper panel}) and the same image after rotation (\textit{lower panel}). The white contour level corresponds to the $3\sigma$ value. \textit{Bottom:}
    The $z$-profile for the artificial galaxy. The horizontal axis is the offset from the galactic plane in arcseconds, while the vertical axis is the logarithm of the flux density in $\mu$Jy/beam. The artificial data is shown in magenta, the fitted disk is shown in dashed line and the halo component corresponds to the dotted line. The total model is plotted in bold line.}
    \label{fig:galaxy_fit}
\end{figure}

We developed a software to identify and characterize the presence of extra-planar emission. 
The software splits the galaxy image into multiple strips perpendicular to the galactic disk (in $z$-direction) and averages them together to create mean intensity profiles, hereafter $z$-profiles. The width of this $z$-profile is given by the combined effect of the telescope resolution, the galaxy inclination, the intrinsic height of the galactic plane, and the possible presence of a radio halo. The software models these components by assuming that the galactic plane is an infinitesimally thin disk. When it is inclined and projected onto the sky, its profile along the $z$-direction can be described either by a Gaussian or an exponential function. Each function is convolved with a Gaussian-shaped beam to model the effect introduced by the telescope resolution. If a single function is not enough to fit the $z$-profile, a second exponential component is added to model the emission from the halo. 

We tested the software using artificial galaxies. These were created considering different combinations of disk diameters and heights, besides an inclination range between $80^{\circ}$ and $90^{\circ}$. Our software could find the presence of the halos for artificial galaxies with inclinations greater than $\sim83^{\circ}$. We found no dependence of the halo scale heights with the inclination. In Fig.~\ref{fig:galaxy_fit}, we show the results obtained for an artificial galaxy created with an inclination of $84^{\circ}$ and a halo with 10 arcseconds FWHM. The best fit was found for a two-component function with an exponential disk, properly identifying the halo.\\

\section{Current status}
\label{sec:current}

\begin{figure}
    \includegraphics[width=0.4\textwidth]{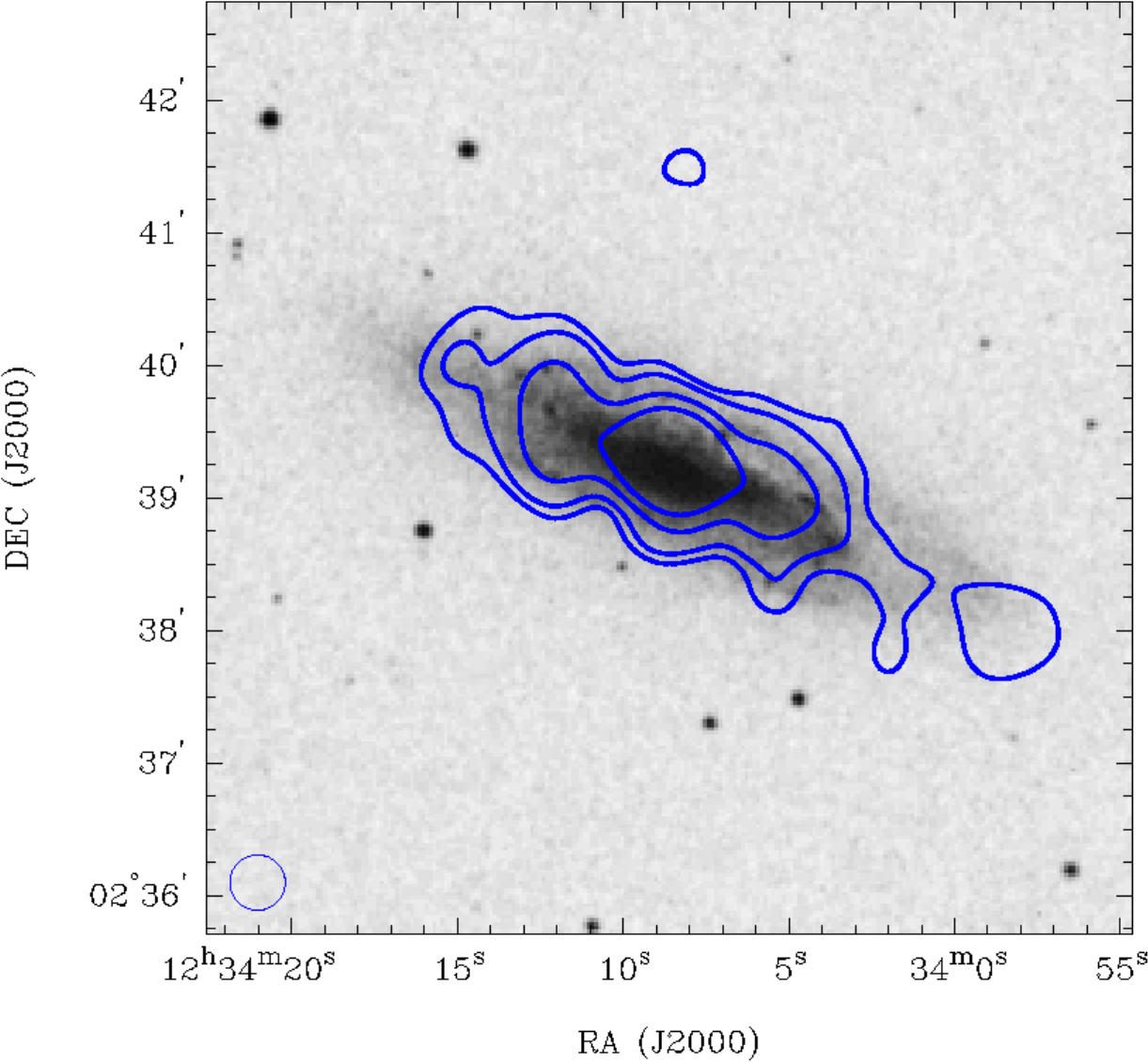}
    \caption{NGC\,4527 galaxy. The radio contour map at 550~MHz is superimposed on a DSS image. The contour levels are 3.5, 5, 10, 20 in units of $\sigma$(=1~mJy\,beam$^{-1}$). The synthesised beam of $25'' \times 25''$ is shown at the bottom-left corner.}
    \label{fig:galaxy}
\end{figure}

We found that the software we developed to detect and characterize the extra-planar emission works correctly by testing it using the artificial galaxies described in Sect. \ref{sec:meth}. The three galaxies of the sample were observed with the Giant Metrewave Radio Telescope (GMRT) at band-4 (550~MHz) and band-5 (1420~MHz). In order to attain 10$\mu$Jy beam$^{-1}$ rms and considering a fudge factor of 2.5 following ETC recommendations, we observed each source for 3 and 3.5 hours, at bands 5 and 4 respectively, including overhead. Because NGC\,5690 is dimmer and the smallest, this galaxy was observed during 4 and 3.5 hours at bands 5 and 4. Currently, we are calibrating and processing the data. A very preliminary image of the galaxy NGC\,4527 with synthesised beam of $25'' \times 25''$ and a noise level of 1~mJy\,beam$^{-1}$, is shown in Figure~\ref{fig:galaxy}. \\

\vspace{0.3cm}

{\bf Acknowledgements:}  GER acknowledges financial support from the State Agency for Research of the Spanish Ministry of Science and Innovation
under grants PID2019-105510GB-C31AEI/10.13039/501100011033/ and
PID2022-136828NB-C41/AEI/10.13039/501100011033/, and by``ERDF A way of
making Europ'', by the ``European Unio'', and through the ``Unit of
Excellence Mar\'ia de Maeztu 2020-2023'' award to the Institute of
Cosmos Sciences (CEX2019-000918-M). Additional support came from PIP 0554 (CONICET).

\end{document}